\def\pndate{25 November 1997}

\documentstyle[aps,preprint,prl]{revtex}
\def\etal{{\it et al.}}
\def\half{{\textstyle{1\over2}}} %puts a small half in a displayed eqn

\def\ex#1{{\rm e}^{#1}}                 % exponential
\def\eg{{\em e.g.}}
\def\eff{_{\rm eff}}
\def\dd{{\rm d}}

\def\kbt{k_{\rm B}T}

\def\CO{{\cal O}}
\def\CZ{{\cal Z}}
\def\CW{{\cal W}}
\def\rmi{{\rm i}}

\def\davg#1{[\![#1]\!]}
\def\elas{_{\rm elas}}
\def\ss#1{\bbox{\sf #1}}\def\ssscript#1{\bbox{\sf #1}}
\def\bOmega{\bbox{\Omega}}

\begin{document}
\preprint{UPR-781T}

\title{Sequence Effects on DNA Entropic Elasticity}
\author{Philip Nelson}
\address{Department of Physics and Astronomy,
University of Pennsylvania\\Philadelphia, PA  19104 USA}
\date{\pndate}
\maketitle

\begin{abstract}
DNA stretching experiments are usually interpreted using the worm-like
chain model; the persistence length $A$ appearing in the model is then
interpreted as the elastic stiffness of the double helix.
In fact the persistence length obtained by this method is a
combination of bend stiffness and {\sl intrinsic bend} effects
reflecting sequence information, just as at zero stretching
force. This observation resolves the discrepancy between the value of
$A$ measured in these experiments and the larger
``dynamic
persistence length'' measured by other means.
On the other hand, the {\sl twist} persistence length deduced
from torsionally-constrained stretching experiments suffers no such
correction. Our calculation is very simple and analytic; it applies to
DNA and other polymers with weak intrinsic disorder.

\end{abstract}

\noindent{\sl Introduction and Summary: }
The DNA in living cells is often described as a passive database
of pure information, the genome. In fact, however, the DNA
molecule itself actively collaborates in its own packaging,
transcription, regulation, and repair~\cite{albe89a}. Unraveling
the underlying mechanisms of these crucial processes requires an
understanding of the basic mechanical properties of the DNA
duplex. For example, the fundamental unit of DNA packaging, the
nucleosome, is  delicately balanced between elastic
stresses and bonding energies~\cite{pola95a}; an accurate account of
the former is clearly important for analyzing the stability of the
whole complex. Since nucleosomal DNA is under torsional as well as
bending stress~\cite{luge97a}, an accurate model incorporating both
twist and bend is needed.

Recently a new class of experiments has permitted precise physical
control over {\it single molecules} of DNA~\cite{aust97a}. For
example, a single molecule of known contour length $L$ can be subjected to
known stretching force $f$ at its ends and the resulting extension
(end-to-end length) $Z$ measured. Simple arguments from polymer
physics then predict that $Z<L$ since thermal fluctuations keep a
flexible rod from being perfectly straight; $Z$ approaches $L$ at
large $f$. Remarkably, Bustamante \etal{} found that a very
simple model, the ``worm-like chain'', fit the force-extension
data over four orders of magnitude in $f$ \cite{bust94a}. The model
attributes to
DNA just one parameter, the bend persistence length $A\eff$;
subsequent experiments have refined its value to%
%\footnote{The authors
%of \cite{wang97a} separated the true elastic contribution to $A$ from
%the electrostatic contribution by extrapolating to high salt. They
%found the more usual value  $A\eff\approx47\,$nm in physiological salt
%concentrations.}
 \cite{fn1} $A\eff=40\,$nm \cite{wang97a}.
In a refinement of the
technique, Strick \etal{} devised a {\it torsionally-constrained}
stretching experiment~\cite{stri96a,stri97a}; analyses of the
corresponding directed walk problem led to values of the twist
persistence length $C\eff$ between  75 and
120~nm~\cite{mark97b,bouc97a,moro97a,moro97b}; in each case
$C\eff/A\eff$ was found to exceed unity.

The purpose of this note is to show that the value $A\eff$ measured by
stretching experiments does {\it not} directly
reflect the bend stiffness of the DNA helix, but rather a certain
combination of stiffness and {\it disorder} induced by the
sequence of natural DNA. Since these two effects will enter in
different combinations in other circumstances, for example the
nucleosome binding energy, it is important to disentangle them. In
fact $A\eff$ {\it underestimates} the true elastic
stiffness $A$, while $C\eff$ accurately reflects the true $C$, as
announced in~\cite{moro97a}. Thus the large observed value of
$C\eff/A\eff$ is perhaps not as mysterious as it at first seems.

Recently Bensimon \etal{} have independently studied these and
other issues~\cite{bens97a}. Using a different model from ours,
they found analytical formul\ae{} for
low-force stretching and numerical results for all $f$, at both strong
and weak disorder. Below we will restrict to the case of weak
disorder, the case relevant for DNA. In this limit
the calculation becomes very simple. The result obtained here for
$A\eff$ differs from~\cite{bens97a}, as  described below.
We will also retain the torsional degree of freedom needed to study
the twist stiffness.

The result of this note is perhaps not surprising in the light of
extensive earlier work on DNA coils at {\it zero} applied tension.
A uniform rigid stack of monomers must form some sort of helix,
and in particular such a helix will have a straight axis in its
undeformed state. DNA, however, is a stack of four different types
of unit. The sequence of natural DNA has a small component with
period equal to the helix repeat~\cite{trif80a}, but mainly the
sequence imparts random natural bends to the rod \cite{wido96a}. Trifonov
\etal{} noted that even in the absence of any thermal fluctuations
a randomly-kinked rod would follow a random walk of some
persistence length $P$, which they called the ``static persistence
length.'' They
argued that the effective persistence length of such a
coil at nonzero temperature would be~\cite{trif87a}
\begin{equation}
A\eff=A/(1+\lambda)\ ,
\label{e1}\end{equation}
where $A\cdot\kbt$ is the true elastic stiffness of the rod and
$\lambda\equiv A/P$, and they verified formula~(\ref{e1}) with Monte Carlo
simulations%
%.\footnote{Later analytical calculations also supported the
%formula~\cite{sche95a}, though Bensimon \etal{} have criticized
%Schellman and Harvey's treatment of the disorder \cite{bens97a}. }
 \cite{fn2}. Trifonov \etal{} computed
the numerical value $P=216\,$nm and hence $\lambda=0.3$ starting from sequence
information and estimates of the wedge angles. Later Bednar
\etal{} measured $\lambda$ more directly by comparing random coils
of natural DNA to synthetic constructs designed to be straight;
they obtained $A=78\,$nm, $A\eff=45\,$nm, and hence $\lambda=0.4$
\cite{bedn95a,furr97a}%
%.\footnote{Some
%dynamical studies have
%yielded even larger values for $A$, presumably because energy
%barriers make transitions to some conformations very slow,
%effectively freezing them out on short time scales~\cite{song90a}.}
 \cite{fn3}.

One might imagine that under extensional force the kinked rod
would simply follow the usual worm-like chain result with $A$
replaced by $A\eff$ from~(\ref{e1}). Indeed {\sl
this is correct for weak disorder} (small $\lambda$). In
contrast, Bensimon \etal{} found that at weak disorder
$A\eff=A(1-\half\sqrt\lambda)$ \cite{bens97a}, while
Marko and Siggia argued that at high force disorder is immaterial:
$A\eff=A$ \cite{mark95a}.
%, contrary to the conclusion of
%\S??? of ref.~\cite{mark95a}.

For a rod under torsional stress
similarly define $C\eff$ by the torque $\tau$ needed to change
the linking number Lk to a value different from its relaxed value
Lk$_0$:
\begin{equation}
\tau=C\eff\cdot\kbt\cdot \omega_0{{\rm Lk}-{\rm Lk}_0\over
{\rm Lk}_0}\ .
\label{e1.1}\end{equation}
Here $\omega_0=1.85/$nm is the rotation per unit length of relaxed
DNA.  In contrast to~(\ref{e1}) we have
\begin{equation}C\eff=C+\cdots
\label{e2}\end{equation}
where  the ellipsis denotes terms vanishing at high force or
greater than first order in $\lambda$.

\noindent{\sl Calculation: }
We wish to evaluate the extension of a randomly-kinked, flexible
rod under an imposed tension $f$, and later an
applied torque as well. We seek the leading term in an expansion
in weak disorder; the extension to higher orders is
straightforward%
%.\footnote{
%The analysis  neglects the helical character of DNA,
%taking the elasticity as well as the disorder to be isotropic
%about the rod axis. This is a good approximation since the important
%fluctuations are on length scales several times longer than the helical
%repeat; see the Appendix to \cite{moro97b}. We also
%neglect any random variation in the
%elastic constants themselves (see \eg{}~\cite{fuji90a}); sequence
%effects enter only in random preferred angles from each segment to
%the next. We neglect self-avoidance effects, a good approximation at
%moderately high stretching force~\cite{moro97a,moro97b}.
%Finally we will restrict to forces $f\ll1000\,$pN so
%that the extensibility of the DNA duplex itself is
%negligible~\cite{wang97a}.}
 \cite{fn4}.

To describe the rod conformations, let $\hat E_a(s)$ be an
orthonormal triad describing the orientation of the rod segment at
arclength $s$ from the end, with $\hat E_3$ the tangent to the rod
axis. The spatial components $E_{ia}$ of these three vectors thus
form an orthonormal matrix $\ss E(s)$. Let $\bOmega\equiv\ss E^{-1}
\dot{\ss E}\equiv\sum_i \Omega_i\ss T_i$, where the dot denotes
$\dd/\dd s$. $\ss T_i$ are the three antisymmetric $3\times3$
matrices generating rotations, \eg{} $[T_1]_{23}=+1$.  The
elastic energy of a conformation is then:
\begin{equation}
E\elas/\kbt={1\over2}\int_0^L \dd s\,\left[A(
\Omega_1-\zeta_1)^2+A(\Omega_2-\zeta_2)^2+C(\Omega_3-\zeta_3)^2\right]
\ .
\label{e3}\end{equation}
To this energy we now add a term describing the work done by the
external force,
\begin{equation}
-{f\over\kbt}\int \dd s\,E_{33}\ .
\label{e4}\end{equation}

The functions $\zeta_i(s)$ appearing in~(\ref{e3}) specify the random
kinks%
%.\footnote{Eqn.\ (\ref{e3}) is reminiscent of a model of
%membranes with random bends~\cite{mors93a}.
%It differs from the approach of~\cite{bens97a}, where
%each link pivots freely in a cone whose preferred polar angle
%$\theta_0(s)$ was random. As emphasized by Schellman and Harvey, this
%idealization can affect the calculation \cite{sche95a}. Marko and
%Siggia considered a mechanical model of a chain of semicircles of
%radius $P$; this model does not have random curvature at all \cite{mark95a}.
%}
 \cite{fn5}.
We give them an isotropic, Gaussian distribution:
\begin{equation}
\davg{\zeta_i(s)}=0\ ;\quad
\davg{\zeta_i(s)\zeta_j(s')}={\lambda\over A}\delta(s-s')
\left[\begin{array}{ccc}
1&&\\&1&\\&& g
\end{array}\right]_{ij}
\ .
\label{e3.1}\end{equation}
Here the double brackets signify an average over an ensemble of
many possible sequences%
%.\footnote{For intensive quantities such as $Z/L$
%our averaging procedure is equivalent to the actual case of a
%single, very long, sequence.
%%\cut{Also the boundary conditions will be
%%immaterial for the intensive quantities of interest; we will use
%%periodic boundaries for $\hat E_a(s)$ along with fixed linking
%%number.}
%}
\cite{fn6}.
Considering the curve whose curvature is exactly
$\zeta_i(s)$ one can see that $P=A/\lambda$ is the structural
persistence length mentioned above, by calculating
$\davg{\hat E_3(0)\cdot\hat E_3(s)}=1-{s\over P}+\CO(s^2)$.
The constant $g$ in eqn.~(\ref{e3.1}) will drop out of our answers.

Thus even neglecting thermal undulations altogether, straightening
the rod requires some extensional force to overcome the {\it
elastic} energy~(\ref{e3}). We must now introduce {\it entropic}
effects as well, and compute the full extension
\begin{equation}
Z/L=\davg{\langle E_{33}(0)\rangle}\ ,
\label{e3.1.1}\end{equation}
where the angle brackets
are the usual thermal average.

To carry out the calculation, begin with the Euler angle
representation of a rotation matrix, defining three fields
$\theta(s),\ \phi(s),$ and $\psi(s)$ by
\begin{equation}
\ss E=\ex{-\phi{\ssscript T}_3}\ex{-\theta{\ssscript
T}_2}\ex{-\psi{\ssscript T}_3}\ .
\label{e3.2}\end{equation}
To exploit the assumed isotropy  of the rod and its disorder,
define the complex variable $\CW=(\Omega_1+\rmi\Omega_2)/\sqrt2
=\ex{-\rmi\psi}(-\rmi\dot\theta+\dot\phi\sin\theta)/\sqrt2$.
Similarly let $\CZ=(\zeta_1+\rmi\zeta_2)/\sqrt2$, which then obeys
$\davg{\CZ(s)\CZ^*(s')}={\lambda\over A}\delta(s-s')
$ and $
\davg{\CZ(s)\CZ(s')}=0$.
The energy then becomes
\begin{equation}
E\elas/\kbt=\int\dd s\,\left[
A\bigl(|\CW|^2-\CW\CZ^*-\CW^*\CZ\bigr)
+\half C(\dot\psi+\dot\phi\cos\theta+\zeta_3)^2
-{f\over\kbt}\cos\theta
\right]\ .
\label{e5}\end{equation}
We have dropped the divergent constant $\int|\CZ|^2$ from (\ref{e5})
because constants in the energy do not affect thermal averages. It is
now clear that the disorder field $\zeta_3(s)$ may be eliminated from
the last term of (\ref{e5}) by shifting the definition of
$\psi$. Since $\psi$ does not enter the first term, while the next two
terms already contain the disorder field $\CZ$, this shift eliminates
$\zeta_3$ altogether to leading order in the strength $\lambda$. The
physical meaning of this shift is simple. Consider a straight,
isotropic rod with a randomly-rotating reference stripe painted on its
surface. Nothing changes if we pass to a different reference frame
rotated at $s$ by an angle $\int^s\dd s'\,\zeta_3(s')$ relative to the
old one.

What makes our problem interesting is that the disorder $\CZ$ can {\it
not} be so trivially eliminated, due to a clash between the $A$ terms
and the $f$ term. To leading nontrivial order in the disorder strength
$\lambda$ the $\CZ$-terms of (\ref{e5}) contribute
\begin{equation}
1+A^2\int\dd s\dd s'\,
\CW(s)\CZ^*(s)\CW^{*}(s')\CZ(s')\ ,
\label{boltz1}\end{equation}
to the Boltzmann weight $\ex{-E\elas/\kbt}$.
Performing the average
over the $\CZ$ fields eliminates one of the
integrations over $s$, so that the correction factor is the leading
term of $\ex{A\lambda\int\dd s\,|\CW|^2}$. Comparing to (\ref{e5}), we
see that to $\CO(\lambda)$ the effect of disorder is simply to replace
$A$ by  $A\eff=A(1-\lambda)$, leaving $C$ unchanged. This
proves~(\ref{e1},\ref{e2}) since we are working to  first order in
$\lambda$.  To go beyond this order we must be careful to treat the
disorder as quenched, for example via the replica trick
\cite{lube79a}.

We can easily incorporate an external torque $\tau$ applied at
the ends of the rod: $\tau$ couples to the change in Link
density, which in our variables is simply $\dot\psi+\dot\phi+\zeta_3$.
We added $\zeta_3$ to the formula of \cite{fain96a} in order to
measure the {\it change} in Link from the unstressed value; the same
shift in the definition of $\psi$ used earlier thus eliminates
$\zeta_3$ here as well.

Thus within our approximations {\sl the only effect of sequence on
entropic elasticity is to reduce the effective bend persistence
length}, as claimed in eqns.~(\ref{e1},\ref{e2}). The first correction to
eqn.~(\ref{e2}) in powers
of $1/\sqrt f$ is also simple to obtain by substituting
eqn.~(\ref{e1}) into the formula for the effective stiffness given in
\cite{moro97a,moro97b}%
%.\footnote{In fact it is simple to rederive
%the formulas for effective twist stiffness and Link-dependence of the
%extension to leading nontrivial order in $1/\sqrt f$, using
%(\ref{e5}) with $\CZ=\zeta_3=0$. At higher orders, however, this
%method becomes cumbersome and the
%approach of \cite{moro97a,moro97b} is much easier.}
 \cite{fn7}.

\noindent{\sl Discussion: }
The model investigated above may seem highly reductionist, neglecting
as it does all the
specific properties of DNA, \eg{} the specific bends at particular base-pair
junctions. Indeed we have used a continuum model, where there are no
base-pairs at all. But it is precisely the existence of a good continuum
limit, despite the very singular form of the assumed
disorder~(\ref{e3.1}),  which gives the result universality. Like the
phenomenon of entropic elasticity itself,  random kinks affect the
force-extension curve via fluctuations over length scales much longer
than a base-pair.

The analysis given here explains the qualitative
success of models without disorder in fitting DNA stretching
experiments. It also predicts that single-molecule stretching
experiments on long,
intrinsically-straight DNA would show the same increase in effective
persistence length seen at zero force, for example in
\cite{bedn95a,furr97a}.  More importantly, it implies that the elastic
stiffness relevant for deformation of a given segment of DNA on
scales shorter than a micron is considerably greater than the value
obtained by fitting the worm-like chain model to stretching experiments.

\acknowledgments
I would like to thank
J.~M.~Schurr for a question which led to this work, and R.~D.~Kamien
and T.~C.~Lubensky for valuable discussions.
This work was supported in part by  NSF grant DMR95--07366.

\newpage
\bibliographystyle{prsty}
%\bibliography{dna,bends}

\begin{thebibliography}{10}

\bibitem{albe89a}
B. Alberts {\it et~al.}, {\em Molecular biology of the cell} (Garland, New
  York, 1989).

\bibitem{pola95a}
K.~J. Polach and J. Widom, J. Mol. Biol. {\bf 254},  130  (1995).

\bibitem{luge97a}
K. Luger {\it et~al.}, Nature {\bf 389},  251  (1997).

\bibitem{aust97a}
R.~H. Austin {\it et~al.}, Physics Today {\bf 50},  32  (1997).

\bibitem{bust94a}
C. Bustamante, J.~F. Marko, E.~D. Siggia, and S. Smith, Science {\bf 265},
  1599  (1994).

\bibitem{fn1}
The authors of \cite{wang97a} separated the true elastic contribution to $A$
  from the electrostatic contribution by extrapolating to high salt. They found
  the more usual value $A\eff\approx47\,$nm in physiological salt
  concentrations.

\bibitem{wang97a}
M.~D. Wang {\it et~al.}, Biophys. J. {\bf 72},  1335  (1997).

\bibitem{stri96a}
T. Strick {\it et~al.}, Science {\bf 271},  1835  (1996).

\bibitem{stri97a}
T.~R. Strick, J.-F. Allemand, D. Bensimon, and V. Croquette, preprint  (1997).

\bibitem{mark97b}
J.~F. Marko and A. Vologodskii, Biophys. J. {\bf 73},  123  (1997).

\bibitem{bouc97a}
C. Bouchiat and M. M\'ezard, preprint {\tt cond-mat/9706050}  (1997).

\bibitem{moro97a}
J.~D. Moroz and P. Nelson, Proc. Natl. Acad. Sci. USA {\bf in press},
  (1997).

\bibitem{moro97b}
J.~D. Moroz and P. Nelson, preprint  (1997).

\bibitem{bens97a}
D. Bensimon, D. Dohmi, and M. M\'ezard, preprint {\tt cond-mat/9711051}
  (1997).

\bibitem{trif80a}
E. Trifonov and J. Sussman, Proc. Natl. Acad. Sci. USA {\bf 77},  3816  (1980).

\bibitem{wido96a}
J. Widom, J. Mol. Biol. {\bf 259},  579  (1996).

\bibitem{trif87a}
E.~N. Trifonov, R.~K.-Z. Tan, and S.~C. Harvey,  in {\em {DNA} bending and
  curvature}, edited by W.~K. Olson, M.~H. Sarma, and M. Sundaralingam (Adenine
  Press, Schenectady, 1987), pp.\ 243--254.

\bibitem{fn2}
Later analytical calculations also supported the formula (J. A. Schellman and
  S. C. Harvey, Biophys. Chem. {\bf55}, 95 (1995)), though Bensimon \etal{}
  have criticized Schellman and Harvey's treatment of the disorder
  \cite{bens97a}.

\bibitem{bedn95a}
J. Bednar {\it et~al.}, J. Mol. Biol. {\bf 254},  579  (1995).

\bibitem{furr97a}
P. Furrer {\it et~al.}, J. Mol. Biol. {\bf 266},  711  (1997).

\bibitem{fn3}
Some dynamical studies have yielded even larger values for $A$, presumably
  because energy barriers make transitions to some conformations very slow,
  effectively freezing them out on short time scales (L. Song and J.M. Schurr,
  Biopolymers {\bf30}, 229 (1990)).

\bibitem{mark95a}
J. Marko and E. Siggia, Macromolecules {\bf 28},  8759  (1995).

\bibitem{fn4}
The analysis neglects the helical character of DNA, taking the elasticity as
  well as the disorder to be isotropic about the rod axis. This is a good
  approximation since the important fluctuations are on length scales several
  times longer than the helical repeat; see the Appendix to \cite{moro97b}. We
  also neglect any random variation in the elastic constants themselves (see
  \eg{} B. S. Fujimoto and J. M. Schurr, Nature {\bf344}, 175 (1990)); sequence
  effects enter only in random preferred angles from each segment to the next.
  We neglect self-avoidance effects, a good approximation at moderately high
  stretching force~\cite{moro97a,moro97b}. Finally we will restrict to forces
  $f\ll1000\,$pN so that the extensibility of the DNA duplex itself is
  negligible~\cite{wang97a}.

\bibitem{fn5}
Eqn.\ (\ref{e3}) is reminiscent of a model of membranes with random bends (D.
  C. Morse and T. C. Lubensky, J. Phys. II (France) {\bf3}, 531 (1993)). It
  differs from the approach of~\cite{bens97a}, where each link pivots freely in
  a cone whose preferred polar angle $\theta_0(s)$ was random. As emphasized by
  Schellman and Harvey, this idealization can affect the calculation
  \cite{fn2}. Marko and Siggia considered a mechanical model of a chain of
  semicircles of radius $P$; this model does not have random curvature at all
  \cite{mark95a}.

\bibitem{fn6}
For intensive quantities such as $Z/L$ our averaging procedure is equivalent to
  the actual case of a single, very long, sequence.

\bibitem{lube79a}
T.~C. Lubensky,  in {\em Ill-condensed matter}, edited by R. Balian, R.
  Maynard, and G. Toulouse (Elsevier, New York, 1979).

\bibitem{fain96a}
B. Fain, J. Rudnick, and S. \"{O}stlund, Phys. Rev. {\bf E55},  7364  (1996).

\bibitem{fn7}
In fact it is simple to rederive the formulas for effective twist stiffness and
  Link-dependence of the extension to leading nontrivial order in $1/\sqrt f$,
  using (\ref{e5}) with $\CZ=\zeta_3=0$. At higher orders, however, this method
  becomes cumbersome and the approach of \cite{moro97a,moro97b} is much easier.

\end{thebibliography}
\newcommand{\noopsort}[1]{} \newcommand{\printfirst}[2]{#1}
  \newcommand{\singleletter}[1]{#1} \newcommand{\switchargs}[2]{#2#1}

\end{document}